\documentclass[twocolumn,showpacs,preprintnumbers,amsmath,amssymb,prl]{revtex4}

\usepackage{graphicx}
\usepackage{dcolumn}
\usepackage{bm}
\usepackage{multirow} 

\usepackage[latin1]{inputenc}
\usepackage[T1]{fontenc}
\usepackage{textcomp}
\usepackage{units}
\usepackage{xspace}

\usepackage[novolumeabbr]{unitsdef} 
\usepackage{amsmath,amssymb,bm} 

\newlength{\fntxvi} 
\newcommand{\chem}[1]
{{\fontencoding{OMS}\fontfamily{cmsy}\selectfont
  \fntxvi\the\fontdimen16\font
  \fontdimen16\font=3pt\fontdimen17\font=3pt
  $\mathrm{#1}$
  \fontencoding{OMS}\fontfamily{cmsy}\selectfont
  \fontdimen16\font=\fntxvi 
  }}

\newcommand{\chemcap}[1]  
{{\fontencoding{OMS}\fontfamily{cmsy}\selectfont
  \fntxvi\the\fontdimen16\font
  \fontdimen16\font=3pt
  $\mathrm{#1}$
  \fontencoding{OMS}\fontfamily{cmsy}\selectfont
  \fontdimen16\font=\fntxvi}}

\newcommand{\chemsec}[1]  
{$\mathrm{#1}$}

\newcommand{\dif}[0]{\mathrm{d}}

\newcommand{\excl}[1]{} 
\newcommand{\LiLCOformula}[0]{\ensuremath{\mathrm{La}_2\mathrm{Cu}_{1-x}\mathrm{Li}_x\mathrm{O}_4}}
\newcommand{\LSCOformula}[0]{\ensuremath{\mathrm{La}_{2-x}\mathrm{Sr}_x\mathrm{CuO}_4}}
\newcommand{\LiLCO}[0]{LLCO}
\newcommand{\LSCO}[0]{LSCO}

\newcommand{\imperm}[0]{\ensuremath{\varepsilon''}}
\newcommand{\reperm}[0]{\ensuremath{\varepsilon'}}

\newunit{\millikelvin}{\milli\kelvin}
\newunit{\faradpermeter}{\farad\unittimes\meter\unitsuperscript{-1}}

\begin{document}

\preprint{APS/123-QED}

\title{Measurement of low energy charge correlations in underdoped spin-glass La-based cuprates using impedance spectroscopy}

\author{G.~R.~Jelbert$^1$, T.~Sasagawa$^2$, J.~D.~Fletcher$^3$, T.~Park$^{4,5}$, J.~D.~Thompson$^4$, and C.~Panagopoulos$^{1,6}$}
\affiliation{$^1$Cavendish Laboratory, University of Cambridge, Cambridge CB3 0HE, UK\\
$^2$Materials~and~Structures~Laboratory, Tokyo~Institute~of~Technology, Kanagawa 226-8503, Japan\\
$^3$H.~H.~Wills Physics Laboratory, University of Bristol, Bristol BS8 1TL, UK\\
$^4$Los Alamos National Laboratory, Los Alamos, New Mexico 87545, USA\\
$^5$Department of Physics, Sungkyunkwan University, Suwon 440-746, Korea\\
$^6$Department of Physics, University of Crete and FORTH, 71003 Heraklion, Greece}

\date{\today}

\begin{abstract}
We report on the charge kinetics of \chem{La_2CuO_4} lightly doped with Li and Sr.  Impedance spectroscopy measurements down to 25\millikelvin and from 20\hertz to 500\kilohertz reveal evidence for low energy charge dynamics, which slow down with decreasing temperature.  Both systems are acutely sensitive to stoichiometry.  In the case of Sr substitution, which at higher carrier concentration evolves to a high temperature superconductor, the ground state in the pseudogap-doping regime is one of spatially segregated, dynamic charge domains.   The charge carriers slow down at substantially lower temperatures than their spin counterparts and the dynamics are particularly sensitive to crystallographic direction.  This is contrasted with the case of Li-doping.
\end{abstract}

\maketitle

 
Hole-doping into the \chem{CuO_2} planes of the antiferromagnetic Mott insulating compound, \chem{La_2CuO_4}, can be achieved for example with Sr or Li.  In the case of \LiLCOformula\ (\LiLCO), the holes tend to be localized near the Li-sites \cite{sarrao01}, whereas with \LSCOformula\ (\LSCO) they are more mobile \cite{sachdev01,kivelson01}.  In both systems doping ($x$) suppresses the N\'eel temperature ($T_N$) and a short-range ``spin-glass'' phase is exposed near $x=0.03$ at $\sim8$\kelvin \cite{kastner01,chou01,sarrao01,niedermayer01,julien01,wakimoto01,lavrov01,panagopoulos03,sasagawa01,singer01,luscher01}. It is only for $x\geq 0.055$ that superconductivity emerges in \LSCO.  \LiLCO, on the other hand, remains an insulator \cite{kastner01}, yet has a similar magnetic phase diagram at low dopings \cite{sasagawa01,wakimoto01}.  The intermediate spin-glass region in \LSCO\ allows the study of spin and charge dynamics prior to the onset of high temperature superconductivity (HTS).  A wide range of investigations in this intermediate region have provided evidence for the presense of antiferromagnetic domains separated by anti-phase boundaries \cite{lavrov01,wakimoto01,matsuda01}.  In so far as the charge is concerned, it remains unclear whether the system is conducting at these dopings.  There have been indications of spatial segregation of charge into locally ordered regions \cite{julien01,ando03,ando04,kohsaka01,dumm01,tassini01,sugai01}.  Recent infrared studies are inconsistent with the notion of static charge ordering \cite{padilla01}.  These results are supported by low frequency resistance noise measurements \cite{raicevic01,raicevic02}.  

To elucidate the charge dynamics we performed low frequency impedance spectroscopy (20\hertz to 500\kilohertz) on \LiLCO\ and \LSCO, both with $x=0.03$ --- a probe sensitive to correlated domains \cite{kremer01}.  Our studies show that the charge slows down with decreasing temperature ($T$) in both systems.  We observe a dispersion in the conductivity that we attribute to dielectric losses in both systems.  We show that \LSCO\ in the pseudogap-doping regime studied here manifests spatial segregation of holes into locally ordered, low energy, dynamic regions. The charge carriers slow at temperatures a couple of orders of magnitude lower than their spin counterparts --- an anomalous situation in glassy systems in general.  Significantly these low energy charge dynamics occur primarily along the \chem{CuO_2} planes; and none of these is evidenced in \LiLCO, which does not superconduct at any doping. 

The \LiLCO\  and \LSCO\ ($x=0.03$) high-quality single crystals were grown using the traveling-solvent floating-zone method \cite{sasagawa02,sasagawa01}.  In order to eliminate possible hole doping by excess oxygen, the crystals were carefully heat treated under reducing conditions.  The lithium concentrations were estimated to within $\pm 0.003$ \cite{sasagawa01}.  The crystal axes were determined by the x-ray Laue backscattering technique and samples were cut with a wire saw for both in-plane and out-of-plane measurements.  Different contacts were used to ensure that the results were intrinsic, including evaporated Au and cured Dupont 6838 Ag paste.  We tried depositing an \chem{Al_2O_3} layer and subtracting its contribution, but the results were unchanged.  We also varied the geometry to confirm that the effects were not from depletion layers at the contacts.  We performed our measurements down to $1.3\kelvin$ in He dewars and to 300\millikelvin using an adiabatic demagnetisation refrigerator.  Measurements were extended to 25\millikelvin using two different dilution refrigerators.  We measured the impedance, $Z$ and the phase angle, $\theta$. The phase angle varied from approximately $0\degree$ (resistive) at room temperature and low frequency ($f\approx100\hertz$) to $\lesssim 90\degree$ (low-loss dielectric) at low temperatures and higher frequency ($f\approx100$\kilohertz).  The impedance varied by up to nine orders of magnitude in the course of an experiment.  To compare such extreme variance we extracted from our impedance measurements the real $\reperm = -\frac{d\sin\theta}{\omega Z \varepsilon_0 A}$ and imaginary permittivity $\imperm = \frac{d\cos\theta}{\omega Z \varepsilon_0 A}$, and the conductivity $\sigma = \omega\varepsilon_0\imperm$, where $d$ is the distance between the contacts, $A$ is the area of the contacts, $\omega=2\pi f$, and $\varepsilon_0=8.854\times10^{-12}$\faradpermeter is the vacuum permittivity \cite{kremer01}.

\begin{figure}[htbp] 
\centering
\includegraphics[bb = 55 170 505 595, clip, width=7cm]{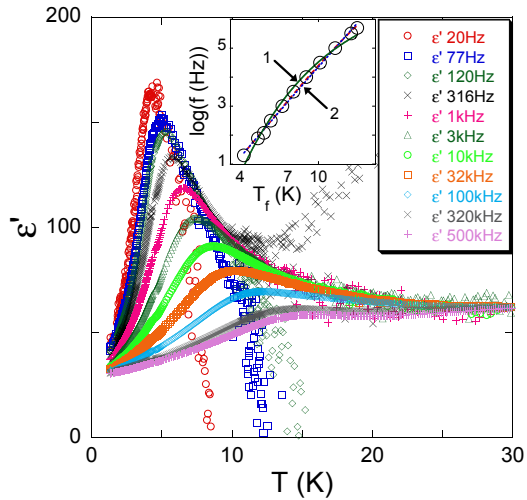}
\caption{ \label{fig:LiLCO-c-re} (Color online) Out-of-plane real permittivity of \LiLCO\ ($x=0.03$) vs $T$.  Inset: $T$ and $f$ dependence of the peaks in \reperm.  (1) Arrhenius fit; (2) Vogel-Fulcher and power-law fits (see text for details). The error-bars (not shown) are smaller than the symbols.}
\end{figure}

\citet{park01} showed that an electronic glass occurs in \LiLCO\ ($x=0.023$), seen by a steplike drop in \reperm\ at a characteristic $T$ which increases with $f$.   Having replicated their results (not shown here), we then studied the effect of additional carrier doping.  A small increase in doping to $x=0.03$ causes an astonishing increase in conductivity and reveals strong peaks in \reperm($T$) (Fig.~\ref{fig:LiLCO-c-re}), the temperature of which increases with $f$.  The roundedness and low frequencies involved suggest polarization of electronic domains with a wide range of characteristic frequencies.  Noise is introduced above the peak $T$ below 1\kilohertz, consistent with domain walls switching in random steps with the application of a slowly varying field.  The effect is less noticeable at higher frequencies where the domains are excited reversibly if at all or at lower temperatures where they are sluggish on these time scales.

The $f$ dependence of the peak in \reperm\ is fitted against three standard forms (Fig.~\ref{fig:LiLCO-c-re}; inset) \cite{kremer01}.  An Arrhenius fit ($\exp(-E/k_BT)$), giving an excitation energy of $E=59\kelvin$, is more convex than the data points.  A Vogel-Fulcher fit ($\exp(-E/k_B(T-T_f))$) gives a freezing temperature of $T_f=-5.8(7)\kelvin<0$ and $E=189(32)\kelvin$.  A power-law fit ($(T-T_f)^n$) gives unphysical parameters ($n=7.5$).  This agrees with the following analysis of the conductivity and will be commented on below.

\begin{figure}[htbp] 
\centering
\includegraphics[bb = 50 185 490 655, clip, width=7cm]{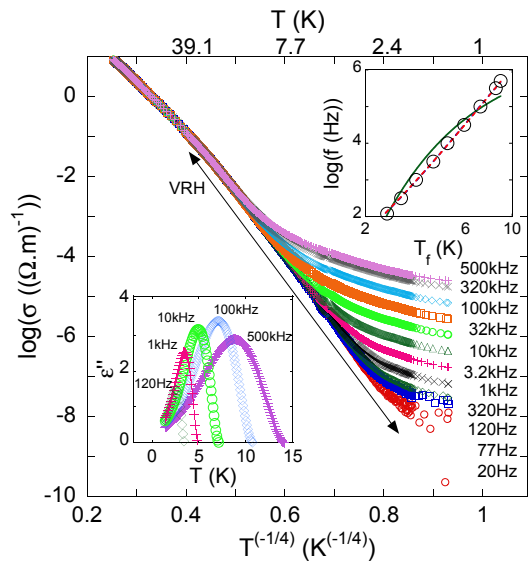}
\caption{ \label{fig:LiLCO-c-VRH} (Color online) Out-of-plane $\sigma$ of \LiLCO\ vs T$^{-1/4}$.  In this plot a straight line represents VRH.  The extrapolated $f$ independent VRH component we subtract from the data is parallel to the double-headed arrow.  Lower inset: representative calculated dielectric losses with VRH component subtracted (see text for details).  Upper inset: $T$ and $f$ dependence of the peaks in the dielectric losses.  The lines are the same fits as in Fig.~\ref{fig:LiLCO-c-re} (inset).  The error-bars (not shown) are smaller than the symbols.}
\end{figure}
  
We extracted the conductivity from our impedance measurements (Fig.~\ref{fig:LiLCO-c-VRH}).  The linearity of the 20\hertz data below $\sim 40\kelvin$ reveals Mott Variable Range Hopping (VRH), $\sigma(T) = \sigma_0\exp(-(T_0/T)^{\beta})$, where $\beta=\frac{1}{(d+1)}$, $d$ is the dimensionality of the hopping \cite{apsley01}. There is a $f$ dependent deviation from VRH.  This may be understood in terms of two components to the losses: VRH (indicated by the double-headed arrow in the main panel of Fig.~\ref{fig:LiLCO-c-VRH}) and dielectric loss.  We subtracted the extrapolated best-fit $f$ independent VRH component from  $\sigma$ to reveal the $f$ dependent loss peaks (Fig.~\ref{fig:LiLCO-c-VRH}; lower inset).  A plot of the $T$ dependence of these peaks (Fig.~\ref{fig:LiLCO-c-VRH}; upper inset) is in reasonable agreement with the data shown in the inset to Fig.~\ref{fig:LiLCO-c-re}.  An Arrhenius fit again fails to describe the data, and a Vogel-Fulcher fit yields $T_f=-6.0(7)\kelvin$ and $E=165(22)\kelvin$.  A power law fit again gives unphysical parameters ($n=7.7$).

The in-plane measurements (not shown) are qualitatively similar, although $\sigma$ and \reperm\ are both an order of magnitude larger than the out-of-plane values at low temperatures.  In the limit of $f\rightarrow0$ our results agree with DC measurements on comparable ceramic samples \cite{kastner01}, while at higher frequencies our experiments reveal similar low energy charge dynamics to those seen in the out-of-plane conductivity above.

The negative $T_f$ values obtained from the Vogel-Fulcher fits suggest that the charge does not freeze in these samples, although it continues to slow as $T$ is reduced.  It is possible that other fits over a lower temperature region will yield a different result: this remains to be checked with further experiments.  Nevertheless, the present results differ to those of \citet{park01} which show $T_f$ of $3-5$\kelvin.  This difference demonstrates the extreme sensitivity of the coupling between the low energy spin and charge dynamics to carrier doping, possibly related to the large $\frac{\dif T_N}{\dif x}(x\sim0.03)$ \cite{sarrao01,sasagawa01}.  At $x=0.03$ none of the standard scaling analyses adequately describe the low energy dynamics of the charge over the temperature range studied.

\begin{figure}[htbp] 
\centering
\includegraphics[bb = 60 20 490 743, clip, width=7cm]{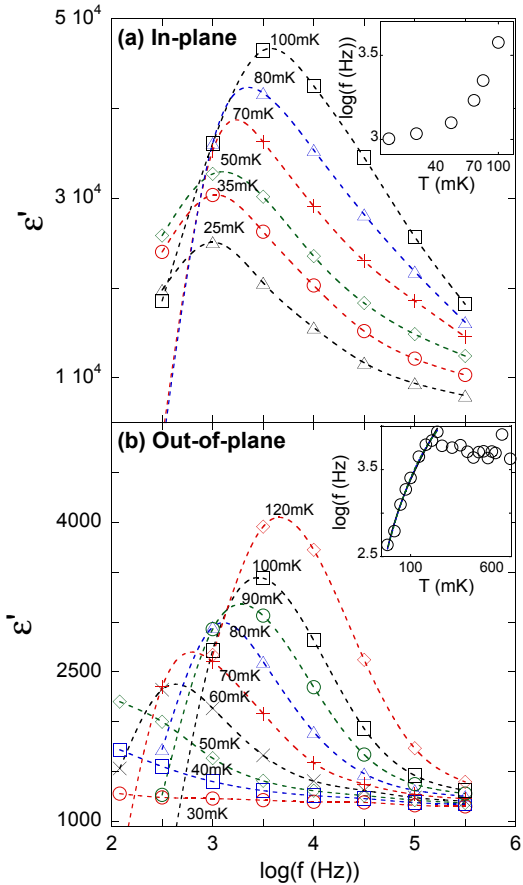}
\caption{ \label{fig:LSCO-re} (Color online) Real permittivity of \LSCO\ vs $f$, with (a) field in-plane and (b) out-of-plane.  The dotted lines are guides to the eye.  Insets are plots of the $T$ and $f$ dependence of the peaks.  The line in the lower inset shows two fits (see text for details). The error-bars (not shown) are smaller than the symbols.}
\end{figure}

We now turn to our results on \LSCO.  Extracting \reperm\ from the impedance measurements reveals broad $T$ dependent resonance-like peaks when plotted against $f$ (Fig.~\ref{fig:LSCO-re}).  The broadening of the dielectric dispersion on cooling can be attributed to a broadening of the distribution of relaxation frequencies.  This is suggestive of a distribution of sizes of electronic domains acting as coherent resonant oscillators. The peak $f$ decreases when $T$ is lowered, indicative of the dynamics slowing down.  These observations are in accord with the emergence of dynamical charge heterogeneities at ultra-low temperatures.  Noise measurements, which reveal distinct switching fluctuations at temperatures of $\lesssim 0.3\kelvin$, further confirm this behavior \cite{raicevic01,raicevic02}.

Although the results with field in-plane and out-of-plane look qualitatively similar, there are some notable differences (Fig.~\ref{fig:LSCO-re}).  The in-plane measurements are an order of magnitude larger.  This may be understood by higher charge carrier mobility in the \chem{CuO_2} planes, enhancing the measured polarization of the domains.  Secondly, the in-plane frequency peaks asymptotically approach $\sim 1$\kilohertz as $T\rightarrow0$ (Fig.~\ref{fig:LSCO-re}(a); inset).  This suggests that the sluggishness of the in-plane domains levels off as $T\rightarrow0$.  The out-of-plane peaks, on the other hand, show a distinct kink at approximately 160\millikelvin (Fig.~\ref{fig:LSCO-re}(b); inset).  We fitted the data below this kink to the three standard forms used previously.  In this instance a Vogel-Fulcher fit reduced to an Arrhenius fit ($T_f\approx 0$), with $E=281\millikelvin$, and a power-law fit yielded $n=1.56(24)$ and $T_f=42(6)\millikelvin$.  Thus, while the in-plane resonant frequency levels off as $T\rightarrow0$, the out-of-plane dynamics continue to slow.  Whether it has finite freezing at $T_f=42\millikelvin$ (as in the power-law fit) or simply activated behavior (as in the Arrhenius fit) remains unclear.

\begin{figure}[htbp] 
\centering
\includegraphics[bb = 65 170 555 645, clip, width=7cm]{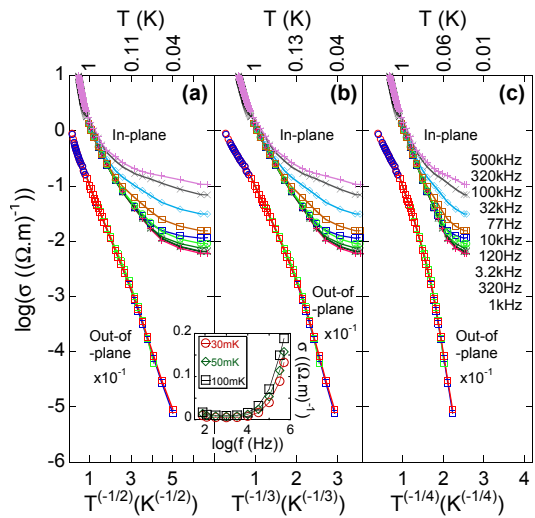}
\caption{ \label{fig:LSCO-VRH} (Color online) $\sigma$ of \LSCO\ vs VRH $T$ exponents, (a) $\beta=1/2$, (b) $\beta=1/3$, and (c) $\beta=1/4$.  Upper curves show in-plane data, lower show out-of-plane, which is shifted down for clarity.  Inset: adjusted in-plane $\sigma$ vs $f$.  The lines are guides to the eye. The error-bars (not shown) are smaller than the symbols.}
\end{figure}

Next we consider the conductivity of this system (Fig.~\ref{fig:LSCO-VRH}).  When the field is in-plane, there is a similar $f$ dependent deviation to that seen in \LiLCO, but at lower temperatures.  This lends credence to the low energy, dynamic charge heterogeneities mentioned above.  We note that although the signature for the onset of dynamic charge heterogeneities occurs in the mK regime (Fig.~\ref{fig:LSCO-re}), the dispersion shown in Fig.~\ref{fig:LSCO-VRH} commences at temperatures of a few kelvin. We again subtracted the VRH component.  The losses are found to be relatively frequency independent for $f\lesssim1$\kilohertz, but highly dispersive for $f\gtrsim1$\kilohertz (Fig.~\ref{fig:LSCO-VRH}; inset).  It is possible to distinguish a similar but significantly smaller effect in the out-of-plane conductivity for $T\lesssim500$\millikelvin (not visible on the scale shown).  The deviation occurs first at the higher frequencies, and for example at 100\millikelvin is monotonic with the 320\kilohertz conductivity 22\% larger than the 120\hertz value.  That this effect is relatively small may be due to the low out-of-plane charge carrier mobility with a concomitant reduction in polarizability.  The observed anisotropy (Figs~\ref{fig:LSCO-re} and \ref{fig:LSCO-VRH}) indicates that the low energy dynamics of the electronic domains occur mainly in the \chem{CuO_2} planes.

We then estimated the barrier energy by fitting the conductivity to the VRH equation.  For the low $T$ in-plane region (Fig.~\ref{fig:LSCO-VRH}(c); $T=1.3-5$\kelvin \& $T=0.04-1$\kelvin) best-fits yield $\beta=1/4$ and $T_0=3.1\times10^3$\kelvin \& $T_0=340$\kelvin.  We contrast these $x=0.03$ results with $x=0.02$ data from the literature. \citet{ellman01} reported $\beta=1/2$ and $T_0=74\kelvin$ for $x=0.02$ at low temperatures (0.3-10\kelvin), and associated the $\beta$ value with weak screening between the interacting charge carriers.  To the best of our knowledge no low $T$ data on $x=0.02$ single crystals is available.  Nevertheless, this comparison shows that a small increase in carrier concentration in this doping region is sufficient to cause an onset of screening of interactions at low temperatures.  This is consistent with a rapid doping induced increase in carrier mobility \cite{komiya01}. In the lowest temperature out-of-plane region (Fig.~\ref{fig:LSCO-VRH}(a); 0.04-0.12\kelvin) we obtain $\beta=1/2$ and $T_0=7\kelvin$, while at higher temperatures (Fig.~\ref{fig:LSCO-VRH}(c); 10-50\kelvin) $\beta=1/4$ and $T_0=163\kelvin$ \cite{note01}.  Between these temperature ranges (Fig.~\ref{fig:LSCO-VRH}(b); 0.5-10\kelvin) a best fit yields $\beta=1/3$ in agreement with other measurements in a similar temperature range \cite{raicevic02}.  Again contrasting with the literature, \citet{birgeneau02} reported $\beta=1/4$ and $T_0=1.1\times10^6$\kelvin for $x=0.02$ in the temperature region 10-100\kelvin.  Thus at higher temperatures we see a sharp reduction in barrier energy as doping increases from $x=0.02$ to $x=0.03$.  We associate the $\beta=1/2$ at low $T$ with weak screening due to lower carrier mobility in this orientation, and the $\beta=1/3$ with a gradual transition between the higher temperature and low temperature regions.

In summary, we performed low $T$, low $f$ impedance spectroscopy to compare the charge kinetics of \LiLCO\ and \LSCO\ ($x=0.03$). Both systems exhibit low energy charge dynamics, which slow down as $T$ is reduced. Comparison with previous work on $x=0.02$ \cite{birgeneau02,ellman01,komiya01,park01}, shows that both systems undergo a sharp increase in mobility as doping increases to $x=0.03$.  Our study of \LSCO\ enabled the identification of a ground state with dynamic charge heterogeneities in the doping range where the pseudogap and spin-glass phase are present.  The correlated slow dynamics identified are largely decoupled from their glassy spin counterparts and lie primarily along the \chem{CuO_2} planes which are responsible for the emergence of HTS at higher charge carrier concentrations.  Notably this was not so for \LiLCO.  Our observations call for a reanalysis of the large region of the HTS phase diagram occupied by a spin-glass phase ($0.01\leq x<0.20$) and the role of low energy charge domains on the emergence of superconductivity in doped Mott insulators.

We thank A.~Carrington, G.~Catalan, D.~Popovi\'c and J.~Scott for discussions, and S.~Goh for technical support.  We acknowledge J.~Sarrao 
for \LiLCO\ ($x=0.023$) crystals used to replicate results \cite{park01}.  G.R.J. is indebted to the Skye Foundation and the Ernest Oppenheimer Memorial Trust for scholarships and I2CAM and Magdalene College for travel grants.  This work is supported by the EPSRC, The Royal Society, the EURYI Scheme and MEXT-CT-2006-039047.  


\begin{thebibliography}{31}
\expandafter\ifx\csname natexlab\endcsname\relax\def\natexlab#1{#1}\fi
\expandafter\ifx\csname bibnamefont\endcsname\relax
  \def\bibnamefont#1{#1}\fi
\expandafter\ifx\csname bibfnamefont\endcsname\relax
  \def\bibfnamefont#1{#1}\fi
\expandafter\ifx\csname citenamefont\endcsname\relax
  \def\citenamefont#1{#1}\fi
\expandafter\ifx\csname url\endcsname\relax
  \def\url#1{\texttt{#1}}\fi
\expandafter\ifx\csname urlprefix\endcsname\relax\def\urlprefix{URL }\fi
\providecommand{\bibinfo}[2]{#2}
\providecommand{\eprint}[2][]{\url{#2}}

\bibitem[{\citenamefont{Sarrao et~al.}(1996)}]{sarrao01}
\bibinfo{author}{\bibfnamefont{J.~L.} \bibnamefont{Sarrao}}
  \bibnamefont{et~al.}, \bibinfo{journal}{Phys. Rev. B}
  \textbf{\bibinfo{volume}{54}}, \bibinfo{pages}{12014} (\bibinfo{year}{1996}).

\bibitem[{\citenamefont{Sachdev}(2003)}]{sachdev01}
\bibinfo{author}{\bibfnamefont{S.}~\bibnamefont{Sachdev}},
  \bibinfo{journal}{Rev. Mod. Phys.} \textbf{\bibinfo{volume}{75}},
  \bibinfo{pages}{913} (\bibinfo{year}{2003}).

\bibitem[{\citenamefont{Kivelson et~al.}(2003)}]{kivelson01}
\bibinfo{author}{\bibfnamefont{S.~A.} \bibnamefont{Kivelson}}
  \bibnamefont{et~al.}, \bibinfo{journal}{Rev. Mod. Phys.}
  \textbf{\bibinfo{volume}{75}}, \bibinfo{pages}{1201} (\bibinfo{year}{2003}).

\bibitem[{\citenamefont{Chou et~al.}(1995)}]{chou01}
\bibinfo{author}{\bibfnamefont{F.~C.} \bibnamefont{Chou}} \bibnamefont{et~al.},
  \bibinfo{journal}{Phys. Rev. Lett.} \textbf{\bibinfo{volume}{75}},
  \bibinfo{pages}{2204} (\bibinfo{year}{1995}).

\bibitem[{\citenamefont{Niedermayer et~al.}(1998)}]{niedermayer01}
\bibinfo{author}{\bibfnamefont{C.}~\bibnamefont{Niedermayer}}
  \bibnamefont{et~al.}, \bibinfo{journal}{Phys. Rev. Lett.}
  \textbf{\bibinfo{volume}{80}}, \bibinfo{pages}{3843} (\bibinfo{year}{1998}).

\bibitem[{\citenamefont{Julien et~al.}(1999)}]{julien01}
\bibinfo{author}{\bibfnamefont{M.-H.} \bibnamefont{Julien}}
  \bibnamefont{et~al.}, \bibinfo{journal}{Phys. Rev. Lett.}
  \textbf{\bibinfo{volume}{83}}, \bibinfo{pages}{604} (\bibinfo{year}{1999}).

\bibitem[{\citenamefont{Wakimoto et~al.}(2000)}]{wakimoto01}
\bibinfo{author}{\bibfnamefont{S.}~\bibnamefont{Wakimoto}}
  \bibnamefont{et~al.}, \bibinfo{journal}{Phys. Rev. B}
  \textbf{\bibinfo{volume}{62}}, \bibinfo{pages}{3547} (\bibinfo{year}{2000}).

\bibitem[{\citenamefont{Lavrov et~al.}(2001)}]{lavrov01}
\bibinfo{author}{\bibfnamefont{A.~N.} \bibnamefont{Lavrov}}
  \bibnamefont{et~al.}, \bibinfo{journal}{Phys. Rev. Lett.}
  \textbf{\bibinfo{volume}{87}}, \bibinfo{pages}{017007}
  (\bibinfo{year}{2001}).

\bibitem[{\citenamefont{Panagopoulos et~al.}(2002)}]{panagopoulos03}
\bibinfo{author}{\bibfnamefont{C.}~\bibnamefont{Panagopoulos}}
  \bibnamefont{et~al.}, \bibinfo{journal}{Phys. Rev. B}
  \textbf{\bibinfo{volume}{66}}, \bibinfo{pages}{064501}
  (\bibinfo{year}{2002}).

\bibitem[{\citenamefont{Singer et~al.}(2002)\citenamefont{Singer, Hunt, and
  Imai}}]{singer01}
\bibinfo{author}{\bibfnamefont{P.~M.} \bibnamefont{Singer}},
  \bibinfo{author}{\bibfnamefont{A.~W.} \bibnamefont{Hunt}}, \bibnamefont{and}
  \bibinfo{author}{\bibfnamefont{T.}~\bibnamefont{Imai}},
  \bibinfo{journal}{Phys. Rev. Lett.} \textbf{\bibinfo{volume}{88}},
  \bibinfo{pages}{047602} (\bibinfo{year}{2002}).

\bibitem[{\citenamefont{L\"uscher et~al.}(2007)\citenamefont{L\"uscher,
  Milstein, and Sushkov}}]{luscher01}
\bibinfo{author}{\bibfnamefont{A.}~\bibnamefont{L\"uscher}},
  \bibinfo{author}{\bibfnamefont{A.~I.} \bibnamefont{Milstein}},
  \bibnamefont{and} \bibinfo{author}{\bibfnamefont{O.~P.}
  \bibnamefont{Sushkov}}, \bibinfo{journal}{Phys. Rev. Lett.}
  \textbf{\bibinfo{volume}{98}}, \bibinfo{pages}{037001}
  (\bibinfo{year}{2007}).

\bibitem[{\citenamefont{Kastner et~al.}(1988)}]{kastner01}
\bibinfo{author}{\bibfnamefont{M.~A.} \bibnamefont{Kastner}}
  \bibnamefont{et~al.}, \bibinfo{journal}{Phys. Rev. B}
  \textbf{\bibinfo{volume}{37}}, \bibinfo{pages}{111} (\bibinfo{year}{1988}).

\bibitem[{\citenamefont{Sasagawa et~al.}(2002)}]{sasagawa01}
\bibinfo{author}{\bibfnamefont{T.}~\bibnamefont{Sasagawa}}
  \bibnamefont{et~al.}, \bibinfo{journal}{Phys. Rev. B}
  \textbf{\bibinfo{volume}{66}}, \bibinfo{pages}{184512}
  (\bibinfo{year}{2002}).

\bibitem[{\citenamefont{Matsuda et~al.}(2002)}]{matsuda01}
\bibinfo{author}{\bibfnamefont{M.}~\bibnamefont{Matsuda}} \bibnamefont{et~al.},
  \bibinfo{journal}{Phys. Rev. B} \textbf{\bibinfo{volume}{65}},
  \bibinfo{pages}{134515} (\bibinfo{year}{2002}).

\bibitem[{\citenamefont{Ando et~al.}(2002)}]{ando03}
\bibinfo{author}{\bibfnamefont{Y.}~\bibnamefont{Ando}} \bibnamefont{et~al.},
  \bibinfo{journal}{Phys. Rev. Lett.} \textbf{\bibinfo{volume}{88}},
  \bibinfo{pages}{137005} (\bibinfo{year}{2002}).

\bibitem[{\citenamefont{Ando et~al.}(2003)\citenamefont{Ando, Lavrov, and
  Komiya}}]{ando04}
\bibinfo{author}{\bibfnamefont{Y.}~\bibnamefont{Ando}},
  \bibinfo{author}{\bibfnamefont{A.~N.} \bibnamefont{Lavrov}},
  \bibnamefont{and} \bibinfo{author}{\bibfnamefont{S.}~\bibnamefont{Komiya}},
  \bibinfo{journal}{Phys. Rev. Lett.} \textbf{\bibinfo{volume}{90}},
  \bibinfo{pages}{247003} (\bibinfo{year}{2003}).

\bibitem[{\citenamefont{Kohsaka et~al.}(2007)}]{kohsaka01}
\bibinfo{author}{\bibfnamefont{Y.}~\bibnamefont{Kohsaka}} \bibnamefont{et~al.},
  \bibinfo{journal}{Science} \textbf{\bibinfo{volume}{315}},
  \bibinfo{pages}{1380} (\bibinfo{year}{2007}).

\bibitem[{\citenamefont{Dumm et~al.}(2003)}]{dumm01}
\bibinfo{author}{\bibfnamefont{M.}~\bibnamefont{Dumm}} \bibnamefont{et~al.},
  \bibinfo{journal}{Phys. Rev. Lett.} \textbf{\bibinfo{volume}{91}},
  \bibinfo{pages}{77004} (\bibinfo{year}{2003}).

\bibitem[{\citenamefont{Tassini et~al.}(2005)}]{tassini01}
\bibinfo{author}{\bibfnamefont{L.}~\bibnamefont{Tassini}} \bibnamefont{et~al.},
  \bibinfo{journal}{Phys. Rev. Lett.} \textbf{\bibinfo{volume}{95}},
  \bibinfo{pages}{117002} (\bibinfo{year}{2005}).

\bibitem[{\citenamefont{Sugai et~al.}(2006)\citenamefont{Sugai, Takayanagi, and
  Hayamizu}}]{sugai01}
\bibinfo{author}{\bibfnamefont{S.}~\bibnamefont{Sugai}},
  \bibinfo{author}{\bibfnamefont{Y.}~\bibnamefont{Takayanagi}},
  \bibnamefont{and} \bibinfo{author}{\bibfnamefont{N.}~\bibnamefont{Hayamizu}},
  \bibinfo{journal}{Phys. Rev. Lett.} \textbf{\bibinfo{volume}{96}},
  \bibinfo{pages}{137003} (\bibinfo{year}{2006}).

\bibitem[{\citenamefont{Padilla et~al.}(2005)}]{padilla01}
\bibinfo{author}{\bibfnamefont{W.~J.} \bibnamefont{Padilla}}
  \bibnamefont{et~al.}, \bibinfo{journal}{Phys. Rev. B}
  \textbf{\bibinfo{volume}{72}}, \bibinfo{pages}{205101}
  (\bibinfo{year}{2005}).

\bibitem[{\citenamefont{Rai\v{c}evi\'c et~al.}(2007)}]{raicevic01}
\bibinfo{author}{\bibfnamefont{I.}~\bibnamefont{Rai\v{c}evi\'c}}
  \bibnamefont{et~al.}, \bibinfo{journal}{Proc. of SPIE}
  \textbf{\bibinfo{volume}{6600}}, \bibinfo{pages}{660020}
  (\bibinfo{year}{2007}).

\bibitem[{\citenamefont{Rai\v{c}evi\'c et~al.}(2008)}]{raicevic02}
\bibinfo{author}{\bibfnamefont{I.}~\bibnamefont{Rai\v{c}evi\'c}}
  \bibnamefont{et~al.} (\bibinfo{year}{2008}), \bibinfo{note}{arXiv:0802.3817}.

\bibitem[{\citenamefont{Kremer and Sch{\"o}nhals}(2003)}]{kremer01}
\bibinfo{author}{\bibfnamefont{F.}~\bibnamefont{Kremer}} \bibnamefont{and}
  \bibinfo{author}{\bibfnamefont{A.}~\bibnamefont{Sch{\"o}nhals}},
  \emph{\bibinfo{title}{Broadband Dielectric Spectroscopy}}
  (\bibinfo{publisher}{Springer}, \bibinfo{year}{2003}).

\bibitem[{\citenamefont{Sasagawa et~al.}(1998)}]{sasagawa02}
\bibinfo{author}{\bibfnamefont{T.}~\bibnamefont{Sasagawa}}
  \bibnamefont{et~al.}, \bibinfo{journal}{Phys. Rev. Lett.}
  \textbf{\bibinfo{volume}{80}}, \bibinfo{pages}{4297} (\bibinfo{year}{1998}).

\bibitem[{\citenamefont{Park et~al.}(2005)}]{park01}
\bibinfo{author}{\bibfnamefont{T.}~\bibnamefont{Park}} \bibnamefont{et~al.},
  \bibinfo{journal}{Phys. Rev. Lett.} \textbf{\bibinfo{volume}{94}},
  \bibinfo{pages}{017002} (\bibinfo{year}{2005}).

\bibitem[{\citenamefont{Apsley and Huges}(1974)}]{apsley01}
\bibinfo{author}{\bibfnamefont{N.}~\bibnamefont{Apsley}} \bibnamefont{and}
  \bibinfo{author}{\bibfnamefont{H.~P.} \bibnamefont{Huges}},
  \bibinfo{journal}{Phil. Mag.} \textbf{\bibinfo{volume}{30}},
  \bibinfo{pages}{963} (\bibinfo{year}{1974}).

\bibitem[{\citenamefont{Ellman et~al.}(1989)}]{ellman01}
\bibinfo{author}{\bibfnamefont{B.}~\bibnamefont{Ellman}} \bibnamefont{et~al.},
  \bibinfo{journal}{Phys. Rev. B} \textbf{\bibinfo{volume}{39}},
  \bibinfo{pages}{9012} (\bibinfo{year}{1989}).

\bibitem[{\citenamefont{Komiya et~al.}(2002)}]{komiya01}
\bibinfo{author}{\bibfnamefont{S.}~\bibnamefont{Komiya}} \bibnamefont{et~al.},
  \bibinfo{journal}{Phys. Rev. B} \textbf{\bibinfo{volume}{65}},
  \bibinfo{pages}{214535} (\bibinfo{year}{2002}).

\bibitem[{not()}]{note01}
\bibinfo{note}{Contrasting the in-plane and out-of-plane values of $T_0$ is
  made difficult by the change in dimensionality. A more effective comparison
  can be made by looking at the slopes of Fig.~\ref{fig:LSCO-VRH}. It will be
  noted that for $T>1$\kelvin, the in-plane conductivity decreases more rapidly
  than the out-of-plane component, and vice versa for $T<1$\kelvin.}

\bibitem[{\citenamefont{Birgeneau et~al.}(1987)}]{birgeneau02}
\bibinfo{author}{\bibfnamefont{R.~J.} \bibnamefont{Birgeneau}}
  \bibnamefont{et~al.}, \bibinfo{journal}{Phys. Rev. Lett.}
  \textbf{\bibinfo{volume}{59}}, \bibinfo{pages}{1329} (\bibinfo{year}{1987}).

\end{thebibliography}

\end{document}